\documentclass[twocolumn,prl,aps]{revtex4}
\usepackage{graphicx}

\newcommand{\be}{\begin{equation}}
\newcommand{\ee}{\end{equation}}
\newcommand{\bea}{\begin{eqnarray}}
\newcommand{\eea}{\end{eqnarray}}
\newcommand{\HH}{{\cal H}}

\newcommand{\p}{\partial}

\newcommand{\la}{\langle}
\newcommand{\ra}{\rangle}
\newcommand{\lb}{\left[}
\newcommand{\rb}{\right]}
\newcommand{\lp}{\left(}
\newcommand{\rp}{\right)}
\renewcommand{\phi}{\varphi}
\renewcommand{\vec}[1]{{\mathbf #1}}

\begin{document}
\title{Dynamical projection of atoms to Feshbach molecules at strong coupling}
\author{R.\,A. Barankov and L.\,S. Levitov}
\affiliation{Department of Physics, Massachusetts Institute of Technology,
77 Massachusetts Ave, Cambridge, MA 02139}

\begin{abstract}
The dynamical atom/molecule projection,
recently used to probe fermion pairing,
is fast compared to collective fermion times, but slow
on the Feshbach resonance width scale.
Theory of detuning-induced dynamics
of molecules coupled to
resonantly associating
atom pairs, employing a time-dependent many-body Green's function
approach, is presented. 
An exact solution is found, predicting a $1/3$ power law
for molecule production efficiency at fast sweep.
The results for $s$- and $p$-wave resonances are obtained and compared.
The predicted
production efficiency agrees with experimental observations
for both condensed
and incoherent molecules away from saturation.
\end{abstract} \pacs{} \keywords{}

\maketitle
\vspace{-10mm}

Cold atomic Fermi gases, magnetically tuned to a Feshbach resonance
region \cite{fermion_expts}, host an intriguing strongly interacting many-body
system \cite{Timmermans01,Holland01,Timmermans99,Griffin02}. Recently, pairing
of fermions near the resonance was probed with the help of dynamical projection
of atomic state on the molecular state \cite{Regal04,Zwierlein04}, achieved by a
sweep
through the resonance, followed by the detection of molecular Bose-Einstein
condensate. The sweep could be made very fast compared to typical fermion time
scales, such as the collision frequency or inverse Fermi bandwidth and pairing
energy gap, making the process a ``snapshot probe'' with regard to the
collective fermion processes.

On a single particle level, however, broad Feshbach resonances studied in
Refs.\,\cite{Regal04,Zwierlein04}, exhibit strong atom-molecule
coupling in a relatively wide detuning range.
In this sense,
the sweep speed \cite{Regal04,Zwierlein04} corresponds to
essentially adiabatic atom/molecule conversion, slow on the scale of the
resonance width. For example, in the JILA experiment \cite{Regal04},
the Feshbach resonance width $\Delta B\simeq 10{\rm\,G}$ translates into
$\Delta\nu\simeq 180{\rm\,MHz}$ in detuning frequency, while
the characteristic time of
the magnetic field sweep, $\tau_*=(dt/d\nu)^{1/2}\simeq
1{\rm\,\mu s}$, is about
$10^2$ times longer than $\Delta\nu^{-1}$.
A similar estimate applies to the MIT fast projection
experiment \cite{Zwierlein04}.
Somewhat paradoxically, the fermions involved in this
``slow'' molecule formation are the same whose
many-body state is being analyzed by the dynamical ``snapshot'' projection.
Thus a correct physical picture of the molecular state swept through the
resonance must combine the adiabatic single-particle and the ``snapshot''
many-body aspects in a seamless way.

Continuing efforts to use atom/molecule projection as investigative tool
call for better understanding of the driven molecular state.
The Landau-Zener model \cite{Mies00,Goral04}, 
which fits the data well near saturation \cite{Hodby04}, focuses
on the adiabatic aspects, ignoring molecule dissociation into
continuous spectrum of atom pairs.
The dynamical mean field
approach \cite{Kokkelmans02,Kohler03,Yurovsky04,Duine04,Haque04,Javanainen04,Dicke_FR,Goral05}, which can be justified for bosons in the atomic BEC
regime, lacks firm foundation in the fermion case. 
Recently, the many-body state
overlap models \cite{Diener04,Avdeenkov04,Perali05,Altman05}
were put forward. While providing some
guidance, these approaches do not account for the experimentally relevant
situation of broad resonance \cite{Regal04,Zwierlein04} when the ``snapshot''
many-body projection is slow on the scale of individual molecule formation.


Our objective is to describe molecules at a sweep
fast compared
to the elastic collisions,
when only the quantum-mechanical
processes are relevant.
We develop a theoretical framework which
accounts for resonance
dissociation/association
in the
presence of time-dependent detuning as well as for
fermion pairing correlations.
We describe the molecules swept through the resonance using a
time-dependent Green's function which fully accounts 
free relative motion of the atoms associating to form molecules.
While our method is quite general and
applicable to Feshbach resonances with any angular momentum, here we focus, for
the sake of concreteness, on the $s$-wave case.
We consider the evolution from equilibrium
at $\nu=\nu_0$, followed by an abrupt linear sweep:
%
\be\label{eq:omega_0(t)}
\nu(t)=\cases{\nu_0, & $t<0$\cr \nu_0-\alpha t, & $t>0$}
\ee
with $\alpha$ the sweep rate.
The generalization
to the $p$-wave
resonances \cite{Ticknor04,Zhang04,Schunck05,Gurarie04,Chevy04}
will be straightforward (see below).

\begin{figure}[t]
\centerline{
\begin{minipage}[t]{3.5in}
\vspace{0pt}
\centering
\includegraphics[width=3.5in]{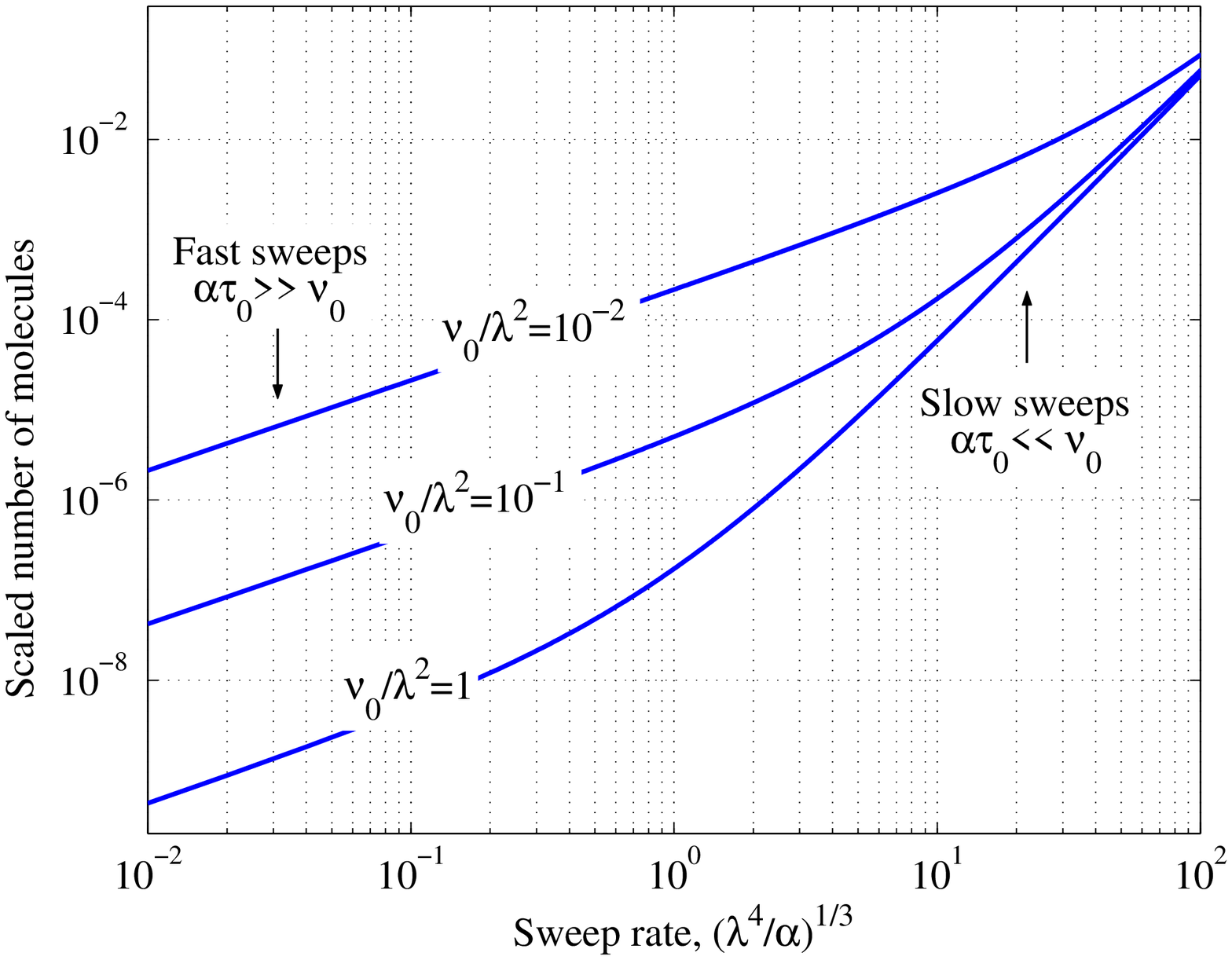}
\end{minipage}
\hspace{-1.75in}
\begin{minipage}[t]{1.6in}
\vspace{1.4in}
\centering
\includegraphics[width=1.5in]{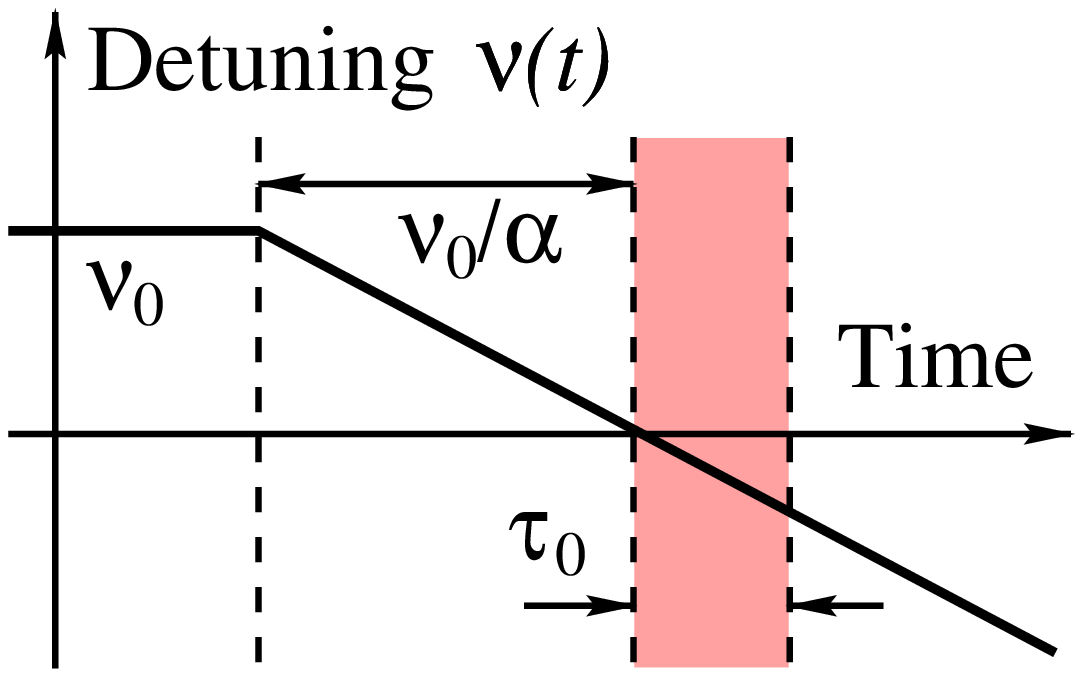}
\end{minipage}
} \vspace{-0.25cm} \caption[]{Molecule number $N_m$ \emph{vs.} the sweep rate
$\lp\lambda^4/\alpha\rp^{1/3}$ at different initial detuning $\nu_0$. The
asymptotic regimes, $N_m\propto \alpha^{-1},\,\alpha^{-1/3}$, correspond to
slow and fast sweep. Inset: Molecular energy
time dependence (\ref{eq:omega_0(t)})
with the time interval corresponding to nonadiabatic evolution marked.
}
\label{fig1}
\end{figure}

Finding the time-dependent Green's function is a nontrivial mathematical
problem, here solved exactly using an idea similar to that of the Wiener-Hopf
method. The important time scale, characterizing the adiabaticity of the
sweep
(see Fig.\,\ref{fig1} inset), is found to be
\be\label{eq:time_scales} \tau_0=\lp \hbar\lambda^2/\alpha^2\rp^{1/3} ,\quad
\lambda=g^2m^{3/2}/4\pi\hbar^3,
\ee
with $g$ the atom-molecule coupling (see Eq.(\ref{H_feshbach})),
and $m$ the atom mass.
The time scale
$\tau_0$ can also be inferred, as noted by Altman and
Vishwanath \cite{Altman05}, from the adiabaticity
condition $\dot\omega\lesssim\omega^2$ for the time-dependent molecule energy
$\hbar\omega$.
Different regimes arise depending on the relation between
$\tau_0$ and $\nu_0/\alpha$, the time it takes the sweep to reach the resonance
(Fig.\,\ref{fig1}). The atom-to-molecule transformation
takes place at times less than $\tau_0$ after crossing the resonance, where
the evolution is nonadiabatic.
At later times, the molecules, dressed by atom pairs,
evolve adiabatically.
For a fast sweep, $\alpha\tau_0\gg\nu_0$, the number of produced molecules
scales with the sweep rate as $\alpha^{-1/3}$, while for slower sweep,
$\alpha\tau_0\ll\nu_0$, the number of molecules scales as $\alpha^{-1}$.

These results agree with the 
molecular number and condensate production efficiency
reported by JILA group \cite{Regal04}.
The sweep speeds $|dt/dB|\approx 10-80{\rm\,\mu s/G}$ \cite{Regal04} correspond
to $\nu_0/\alpha\approx 1-100{\rm\,\mu s}$ with $\nu_0=0.1-1\,{\rm G}$ in the
magnetic field units. The characteristic atom-molecule coupling
$\lambda^2\approx1{\rm\,GHz}$
gives the adiabaticity time $\tau_0\approx 10-20{\rm\,\mu s}$ depending on the
sweep speed. Thus with $0.2<\alpha\tau_0/\nu_0<10$ both the fast and the slow
regimes are realized. Indeed, the molecule number obtained for
different sweep speeds below saturation
(see Fig.\,5 in Ref.\,\cite{Regal04} displaying the data
for $\nu_0=0.12\,{\rm G}$) can be fitted quite accurately with the $1/3$ power law
dependence, $N_m\propto |dt/dB|^{1/3}$, in agreement with our results.
Also reasonable, by the order of magnitude, is the predicted
total number of produced molecules. 
Our conclusions regarding the incoherent molecule
production channel are consistent with the observed independence
of the condensate fraction \cite{Regal04} of the sweep speed.
We obtain the same production efficiency
for 
condensed and incoherent molecules (Eq.(\ref{eq:condensate_fraction})), 
except near saturation.

Let us recall the form of the two-channel Hamiltonian
\cite{Timmermans99}, describing pairs of fermions binding to form molecules at
the resonance:
\be\label{H_feshbach} \HH=\HH_a+\HH_m+ \sum_{\vec p,\vec p'}\lp g b^+_{\vec
p+\vec p'}a_{\vec p\uparrow}a_{\vec p'\downarrow} + {\rm h.c.}\rp
\ee
with $\HH_a= \sum_{\vec p\sigma} \frac{p^2}{2m} a^+_{\vec p\sigma}a_{\vec
p\sigma}$, $\HH_m= \sum_\vec k (\nu+\frac{k^2}{4m}) b_\vec k^+b_\vec k$,
%
and $a_{\vec p\sigma}$, $b_\vec k$ the atom and molecule operators, $\sigma$
the spin ($\hbar=1$).
The detuning $\nu$ is
determined by
the molecule and two-atom Zeeman energy difference, $\nu=\Delta\mu\lp
B-B_0\rp$.

The single molecule Green's function, obtained from Dyson equation
\cite{Bruun04}, has the form
\be\label{eq:mol_green}
G(\omega,k)=\frac{1}{\tilde\omega-\nu-\Sigma(\tilde\omega)}
\,,\quad
\tilde\omega=\omega-\frac{k^2}{4m}+i0
\,,
\ee
where $\Sigma(\omega)=\lambda(-\omega)^{1/2}$ is the self-energy describing
molecule dissociation ($s$-wave),
which arises after integrating over the 3d density of atom pair states
$N(\epsilon)\propto \epsilon^{1/2}$ along with a suitable ultraviolet
regularization \cite{Griffin02}.


For time-independent $\nu$, the molecular state dressed
by atom pairs,
is described by the Green's function pole:
%
\be\label{G,Z}
G_0(\omega)=\frac{Z(\omega)}{\omega-\omega(k)+i0}
\ee
with $\omega(k)$ given by
$\tilde\omega-\Sigma(\tilde\omega)=\nu$.
Near the resonance, at
$
|\nu|\ll \Delta E_\ast=\lambda^2,
$
neglecting $\omega$ compared to $\Sigma(\omega)$, one obtains molecular energy
quadratic in detuning:
\be\label{eq:Mdispersion}
\omega(k)=-(\nu/\lambda)^2+k^2/4m
.
\ee
At $\nu<0$, Eq.(\ref{eq:Mdispersion})
gives the energy of molecules, while at $\nu>0$
it describes a resonance in the two-fermion scattering mediated by 
virtual molecules \cite{virtual_resonance}.
The residue $Z$ defines the bare molecule weight in the physical
molecule state,
$Z^{-1}(\omega)=dG^{-1}/d\omega=1+\frac{\lambda}{2}(-\tilde\omega)^{-1/2}$,
%
which varies from zero to one across the resonance, at $-\Delta E_\ast\lesssim
\nu<0$. At relatively small detuning, $|\nu/\Delta E_\ast|\ll 1$, $Z$
increases linearly: $Z(\omega)\approx 2|\nu|/\lambda^2$.

To investigate molecule formation at the resonance, we consider the Green's
function for the problem with time-dependent detuning $\nu(t)$. In this case,
due to nonlocal character of $\Sigma$ in the time domain, the molecule
evolution is described by an integral-differential equation~\cite{Duine04,Haque04}
\be\label{eq:int-diff}
\lp i\p_t-\nu(t)-{\textstyle \frac{k^2}{4m}}\rp b_k(t)
-\!\int\!\Sigma_k(t,t')b_k(t')dt'=\eta_k(t)
\ee
with $\eta_k(t)=g \int\! e^{-ikx}\psi_\uparrow(x,t)\psi_\downarrow(x,t)
d^3x$ the pairing
field,
and $\psi_\sigma(x,t)=\sum_p a_{p,\sigma} e^{ip x-i\epsilon_p t}$. Here the self-energy is
%
\be\label{eq:Sigma(t,t')}
\Sigma_k(t>t')=\int\! \Sigma_k(\tilde\omega)e^{-i\omega(t-t')}
\frac{d\omega}{2\pi}
=\frac{a e^{-i\frac{k^2}{4m}(t-t')}}{(t-t')^{3/2}}
,
\ee
%
$a=\frac{\lambda}{2\sqrt{i\pi}}$, and $\Sigma(t<t')$ vanishes due to the causality.

The pairing field $\eta$, which acts as a source in
Eq.(\ref{eq:int-diff}), should be taken as a c-number
for the condensed molecules
(with $k=0$), and as an operator for
the incoherent molecules. Generally,
its correlation function
includes
both the coherent and incoherent parts:
\be\label{eq:eta_correlator}
\la \bar\eta_{k,\omega}\eta_{k,\omega}\ra
=(2\pi)^4|\eta_0|^2\delta(\omega-\mu)\delta(k)
+K(\omega,k)
,
\ee
where $\eta_0$ describes a finite amplitude for two fermions
to have opposite momenta in the paired state, with
$\mu\lesssim 2E_F$ the chemical potential of a pair,
and $K(\omega,k)=\la\!\la\bar\eta_{k,\omega}\eta_{k,\omega}\ra\!\ra$
the dynamical pair correlator which is nonzero even
for ideal Fermi gas.
We first consider the coherent molecule
production, treating both $\eta=\eta_0 e^{-i\mu t}$ and $b(t)$ as c-numbers.
The incoherent pair source
$K(\omega,k)$  will be discussed subsequently
below.

The evolution problem (\ref{eq:int-diff}) is non-elementary
due to nonlocality of $\Sigma(t,t')$.
Our approach
employs an idea similar to that used in
the Wiener-Hopf method. We first handle
an auxiliary problem in which the linear
sweep $\nu(t)=\nu_0-\alpha t$
extends from $-\infty$
to $\infty$, and then modify the solution to describe
the situation of interest (\ref{eq:omega_0(t)}).

The auxiliary problem in question is to find $b(t)$, $-\infty<t<\infty$, which
obeys a linear integral-differential equation
\be\label{eq:auxiliary}
\lp \hat\omega-\nu_0-\Sigma(\hat\omega)
+\alpha t\rp b(t)
=\eta(t)
\,,\quad
\hat\omega=i\p_t
\ee
with a source term $\eta(t)$ of a generic form. It is convenient to go to
Fourier representation, in which $t=-i\p_\omega$ and the problem is reduced to
an ordinary differential equation $\lp \omega-\Sigma(\omega)-\nu_0
-i\alpha\p_\omega\rp b(\omega)=\eta(\omega)$ for $b(\omega)=\int e^{i\omega
t}b(t)dt$. Solution of this equation, first order in $\p_\omega$, is found
using the gauge transformation
$b(\omega)\to e^{i\phi(\omega)}b(\omega)$
with the phase $\phi$ satisfying
\be\label{eq:phi'=D0}
-\alpha\phi'(\omega)= D_0(\omega)\equiv \omega-\Sigma(\omega)-\nu_0
.
\ee
This problem is solved by the function
\be\label{eq:b_linear_answer}
b(\omega)=-i\alpha^{-1}e^{i\phi(\omega)}\int_{\omega}^{+\infty}e^{-i\phi(\omega')}\eta(\omega')d\omega'
. \ee
To verify (\ref{eq:b_linear_answer}), one can compare
its behavior to that of $b(\omega)=Ce^{i\phi(\omega)}$,
the solution to the homogeneous problem
(\ref{eq:auxiliary}).
For $\omega$ large and positive, since
$(-\omega-i0)^{1/2}=-i\sqrt{\omega}$, we obtain
the asymptotic behavior
$D_0\approx i\lambda\omega^{1/2}$,
$\phi\approx -i(2\lambda/3\alpha)\omega^{3/2}$,
$e^{\pm i\phi}\propto e^{\pm a\omega^{3/2}}$ ($a=2\lambda/3\alpha$).
Thus $C=0$ is required to eliminate exponential growth.
Indeed, the asymptotic behavior
of the integral in (\ref{eq:b_linear_answer})
at large positive $\omega$ is non-exponential:
$\int_{\omega}^{+\infty}e^{-i\phi(\omega')}\eta(\omega')d\omega'
\approx e^{-i\phi(\omega)}\eta(\omega)/\phi'(\omega)$.
(For any physical source, $\eta(\omega\to\infty)$ is algebraic.)
At the same time, the behavior at large negative $\omega$
does not require special attention:
$\phi$ is real at $\omega<0$, and so the exponentials $e^{\pm i\phi}$
oscillate as  $e^{\pm ia(-\omega)^{3/2}}$
without giving rise
to ``dangerous'' asymptotic behavior.

Now, having found the solution for the sweep spanning the entire range
$-\infty<t<+\infty$, let us consider the sweep trajectory
(\ref{eq:omega_0(t)}). In this case, it is convenient to represent
the function $b(t)$ as a sum $b_<(t)+b_>(t)$, with $b_{>,<}(t)$ nonzero
only at $t\ge 0$ ($t\le 0$), respectively,
obtained by restricting $b(t)$ on the half-line $t\ge 0$ ($t\le 0$).
Then the evolution equation, in operator form written as
$\lp \hat\omega-\nu(t)-\Sigma(\hat\omega)\rp b(t)=\eta(t)$,
can be represented as
\be
\hat D_0 b_<+ (\hat D_0+\alpha \hat t) b_> =
\eta_<+\eta_>
\ee
with $\eta_{>,<}=\theta(\pm t)\eta(t)$ having the same meaning as $b_{>,<}(t)$,
and $\hat D_0\equiv D_0(\hat \omega)$ defined in Eq.(\ref{eq:phi'=D0}).

Let us project the terms on the left hand side on the regions
$t\ge 0$, $t\le 0$, taking into account the constraints due to causality.
The integral operator $\Sigma$ acts only forward, not backward in time,
the property explicit in Eq.(\ref{eq:Sigma(t,t')}).
Thus the function $\lp D_0(\hat\omega)+\alpha \hat t\rp b_>$
is nonzero only at $t>0$, while the function
$D_0(\hat\omega) b_<$ has both the $t>0$ and the $t<0$ parts.
This observation allows to write the problem as two separate
problems for $b_{>,<}(t)$ as follows:
\be
\lb \hat D_0 b_<\rb_<=\eta_<
\,,\quad
( \hat D_0+\alpha \hat t) b_> +\lb \hat D_0 b_<\rb_>=\eta_>
,
\ee
where $[...]_{<,>}$ denotes the part of the function
at $t>0$ ($t<0$), with zero value on the opposite half-line.
Now, we can solve the first equation for $b_<$ and substitute the result in the
second equation, which (after some algebra) can be brought to the form
\be
( \hat D_0+\alpha \hat t) b_> = \hat D_0 \lb \hat D_0^{-1}\eta \rb_>
.
\ee
We note that $b_>$ and the function on the right-hand side
are both nonzero only at $t>0$.
This allows to treat this equation as Eq.(\ref{eq:auxiliary}),
formally extending the
linear time dependence $\alpha t$ to negative $t$. Using the above result,
we obtain the answer in Fourier representation
of the form (\ref{eq:b_linear_answer})
with $\eta$ replaced by
\[
\tilde\eta(\omega)=\hat D_0\lb \hat D_0^{-1}\eta \rb_>
=D_0(\omega)\int \frac{D_0^{-1}(\omega')\eta(\omega')}{\delta-i(\omega-\omega')}\frac{d\omega'}{2\pi}
.
\]
Now, let us consider the source $\eta(t)=\eta_0 e^{-i\mu t}$,
describing coherent fermion pairs with the
chemical potential $\mu/2$ per particle.
In this case, $\eta(\omega)=2\pi\eta_0 \delta(\omega-\mu)$ and
$\tilde\eta(\omega)=D_0^{-1}(\mu)\eta_0 D_0(\omega)/(\delta-i(\omega-\mu))$.
Inserting $\tilde\eta$ in Eq.(\ref{eq:b_linear_answer}),
and using the identity
$(\delta-i(\omega-\mu))^{-1}=\int_0^\infty e^{i(\omega-\mu)\tau}d\tau$,
we find a closed form representation
\bea\label{eq:b0+Delta b}
&& b(\omega)=\frac{A\eta_0 }{\delta-i(\omega-\mu)}+\Delta b(\omega),
\\\nonumber
&& 
\Delta b(\omega)=i A\eta_0  e^{i\phi(\omega)}\int_0^\infty\!\! e^{-i\mu\tau}\tau
\int_{\omega}^{+\infty}\!\!\!\! e^{i\omega'\tau-i\phi(\omega')}d\omega'
d\tau
\eea
with $A=D_0^{-1}(\mu)$. (To obtain (\ref{eq:b0+Delta b}), we transformed
the integral over $\omega'$
by writing $D_0(\omega')
=\alpha d(\omega'\tau-\phi(\omega'))/d\omega'-\alpha\tau$
and integrating by parts.)
%
Since the first term of (\ref{eq:b0+Delta b}) gives the would be
$b(\omega)$ in the absence of the sweep,
$\Delta b(\omega)$
describes the effect of the sweep.

Now, let us analyze the asymptotic behavior of
$b(t)=\int e^{-i\omega t}b(\omega)d\omega/2\pi$ at large positive
$t\gg\tau_0,\nu_0/\alpha$. In this
case, the integral over $\omega$ is controlled by large negative $\omega$,
which can be seen
with the help of the stationary phase
approximation.
Indeed, the saddle point $\omega_\ast$ of $-\omega t+\phi(\omega)$, obtained
from $\phi'=t$, at $t\to+\infty$ implies
$\omega\to-\infty$. With
that in mind, we obtain the asymptotic for $b(t)$ by setting the lower
integration limit
in Eq.(\ref{eq:b0+Delta b}) at
$\omega=-\infty$, leading to the central result of this work:
%
\bea\label{eq:Delta_b}
&&
\Delta b(t)=-\frac{A\eta_0}{2\pi i} F^\ast(t)\int_0^\infty e^{-i\mu\tau}\tau F(\tau)d\tau
,
\\\label{eq:F(t)}
&&
F(t)=\int_{-\infty}^{+\infty}e^{i\omega't-i\phi(\omega')}d\omega'
.
\eea
%
%
%
The qualitative behavior of $F(t)$ can be analyzed using
the stationary phase approximation.
We obtain
$F(t)=\sqrt{-2\pi i/\phi''(\omega_\ast)}e^{i\omega_\ast\tau-i\phi(\omega_\ast)}$,
where the stationary phase equation for $\omega_\ast$, given by
$D_0(\omega_\ast)+\alpha t=0$, has a real solution
$\omega_\ast=-\alpha^2(t-t_0)^2/\lambda^2$
only
for $t>t_0=\nu_0/\alpha$.
Relating the curvature $\phi''$ and the Green's function residue,
$-\alpha\phi''=D_0'=Z^{-1}$, yields the asymptotic form
%
\be\label{eq:Fstationary_phase}
F(t>t_0) = \lp 2\pi i\alpha Z(\omega_\ast)\rp^{1/2}
e^{ -i\frac{\alpha^2}{3\lambda^2}(t-t_0)^3}
\ee
with $Z(\omega_\ast)=2\alpha(t-t_0)/\lambda^2$.
(The self-energy-dominated $D_0(\omega)=-\nu_0-\Sigma(\omega)$,
appropriate for broad $s$-wave
Feshbach resonance,
was used in the
above estimates.)
Thus $F(t)$ grows as $(t-t_0)^{1/2}$
and oscillates at $(t-t_0)/\tau_0 \gg 1$,
decreasing exponentially at $t-t_0<0$.

To apply Eq.(\ref{eq:Delta_b}) to the experimental situation
we take into account that
$\mu\ll\hbar/\tau_0,\,\hbar/t_0$.
(Indeed,
$\mu\lesssim 2E_F$, with
$E_F=0.35\,{\rm \mu K}=50\,{\rm KHz}$ in Ref.\,\cite{Regal04}.)
We evaluate the integral in (\ref{eq:Delta_b}) using the
stationary phase form (\ref{eq:Fstationary_phase}):
%
\be
\Delta b(t)=
(2\alpha Z(t))^{1/2}A\eta_0
e^{ i\frac{\alpha^2}{3\lambda^2}(t-t_0)^3}\!\!
\lp  C_1 t_0+C_2\tau_0\rp
,
\ee
$C_1=(i/3)^{1/2}\Gamma(1/2)$, $C_2=(i/3)^{1/6}\Gamma(5/6)$.
The asymptotic number of molecules at time $t$
is evaluated as
$N_m^{(0)}=|\Delta b(t)|^2$ (see Fig.\ref{fig1}).
The fast and slow sweep regimes can be identified:
at $t_0\ll\tau_0$ we obtain $N_m\propto\alpha^{-1/3}$,
while at $t_0\gg\tau_0$ we have $N_m\propto\alpha^{-1}$.

The incoherent molecule production can be studied in a similar manner. Using
the operator $\eta_k$ as a source
in (\ref{eq:Sigma(t,t')}), and averaging over its dynamical
correlations (\ref{eq:eta_correlator})
we obtain the molecule momentum distribution
%
\be
N_m(k)=\sum_\omega \Big|\frac{A}{2\pi}F^\ast(t)\int\limits_0^\infty\!\! e^{-i\omega\tau}\tau F(\tau)d\tau\Big|^2 K(\omega,k)
.
\ee
with $A=D_0^{-1}(\omega)$.
As a function of frequency, $K$ is nonzero at $\omega\lesssim 2E_F$.
At $E_F\ll\hbar/\tau_0,\,\hbar/t_0$, the expression $|...|^2$ is $\omega$-independent,
as above.
Factoring it out, we conclude that the condensed and incoherent
molecule
production efficiencies are identical.
The molecule condensate fraction is then expressed
through the fermion pair fraction:
\be\label{eq:condensate_fraction}
\frac{N_m^{(0)}}{N_m^{(0)}+\sum_k N_m(k)}=
\frac{|\la\psi_\uparrow(x)\psi_\downarrow(x)\ra|^2}{\la\hat n_\uparrow(x)\hat n_\downarrow(x)\ra}
\ee
$\hat n_\sigma(x)=\bar \psi_\sigma(x)\psi_\sigma(x)$.
We note that the incoherent contribution
exists even in
the absence of pairing. For ideal fermions at density $n$, we have $\sum_\omega
K(\omega,k)=\frac12 g^2n(1-u)^2(1+u/2) \theta(1-u)$, $u=k/2p_F$, which
corresponds to a broad molecule momentum distribution
with $k\le 2p_F$.

The approach presented above yields accurate results for the atom/molecule
projection in a wide range of sweep rates, fast and slow,
as long as the times $\tau_0$, $t_0$ are short on the scale of $E_F$.
The only limitation
stems from the assumption of a steady source,
which describes the situation when
the fraction of atoms converted into molecules is small.
The depletion effects, which are different for the condensed and
incoherent molecules, can be incorporated in the framework
of a quantum kinetic equation.

The above method is applicable to the $p$-wave Feshbach resonance case,
with the essential
modification in the self-energy form
$\Sigma(\omega)\propto (-\omega)^{3/2}$ \cite{Gurarie04,Chevy04}.
This self-energy is an irrelevant perturbation
near the resonance, so we have the atom/molecule conversion time
$\tau_0=\alpha^{-1/2}$ and
$Z=1$, as for weak coupling.
Thus one can use
$D_0(\omega)=\omega-\nu_0$ in (\ref{eq:phi'=D0}),
yielding the result identical to
(\ref{eq:Delta_b}) with
\[F(t)=\int e^{i\omega(t-t_0)+i\omega^2/2\alpha}d\omega=(2\pi i\alpha)^{1/2}
e^{-i\alpha(t-t_0)^2/2}.
\]
%
The number of produced molecules, both condensed
and incoherent, scales as
inverse sweep rate
$\alpha^{-1}$. The production is less efficient than in the
$s$-channel case due to weaker coupling at resonance.

In summary, molecule production at Feshbach resonance
is considered as a many-body problem for which the exact Green's function
is obtained using Wiener-Hopf method.
The theory is applied to the $s$-wave and $p$-wave resonances.
The slow and fast sweep regimes are identified in the $s$-wave case,
controled by the adiabaticity time scale
(\ref{eq:time_scales}).
The predicted power law $1/3$ for the
molecule production, as well as the total molecule number, 
are found to be in agreement
with observations away from saturation \cite{Regal04}.
The independence of the produced condensate fraction on the sweep rate
observed at fast sweep \cite{Regal04}
is also explained by this theory.

We are grateful to Dmitry Petrov for useful comments.

\vspace{-5mm}



\end{document}